# Intelligent Reflecting Surfaces assisted Laser-based Optical Wireless Communication Networks

Ahrar N. Hamad, Walter Zibusiso Ncube, Ahmad Adnan Qidan, Taisir E.H. El-Gorashi and Jaafar M. H. Elmirghani

*Department of Engineering, King's College London, London, United Kingdom,*
*{ahrar.hamad, walter.ncube, ahmad.qidan, taisir.elgorashi, jaafar.elmirghani}@kcl.ac.uk.*

**ABSTRACT**
The increasing demand for wireless networks of higher capacity requires key-enabling technologies. Optical wireless communication (OWC) arises as a complementary technology to radio frequency (RF) systems that can support high aggregate data rates. However, OWC systems face some challenges including beam-blockage. Intelligent reflecting surfaces (IRSs) can offer alternative pathways for the optical signal, ensuring continuous connectivity. In this work, we investigate the potential of using IRS in an indoor OWC network. In particular, we define a system model of indoor OWC that employs IRS in conjunction with angle diversity transmitters (ADT) using vertical-cavity surface-emitting laser (VCSEL) arrays. The VCSEL beam is narrow, directed, and easy to block, however, it can deliver high data rates under eye safety regulations. Simulation results show that the deployment of IRS can significantly improve the achievable data rates of Laser-based OWC systems.

## 1. INTRODUCTION

Optical wireless networks have the potential to deliver ultra-high aggregate data rates, exceeding Gigabits-per-second (Gb/s) barriers, by leveraging the vast and unregulated infrared and visible light spectrums [1]-[6]. However, the modulation bandwidth of traditional light emitting diodes (LED) used in some optical wireless networks limits the high achievable data rates. To address this, laser diodes have emerged as a promising solution to address the limited modulation bandwidth of LEDs, enabling high-speed data transmission in optical wireless communication (OWC) networks [7]-[9]. Specifically, vertical cavity surface emitting lasers (VCSELs) have garnered significant attention as transmitter for short-range communication due to their key advantages [10]-[13]. Moreover, VCSEL properties enable OWC system to potentially achieve high data rates, strong signal integrity, and reduced power consumption [9], [14]. To further enhance the performance and versatility of VCSEL-based OWC systems, angle diversity transmitters (ADT) offer a compelling solution. The use of ADT promises higher capacity, improved coverage, and greater resilience in complex indoor environments. Previous studies have explored the potential of using VCSEL in OWC [14] - [16]. For instance, the authors in [14] present a framework for calculating maximum safe transmit power of single-mode and multi-mode laser beams, ensuring compliance with IEC standards and informing the design of reliable OWC systems. In [15], the laser-based optical wireless network studied by using an array of low-cost VCSELs and zero-forcing to manage multi-user interference. Their findings reveal that wider laser beams and the use of micro-lenses significantly improve the performance of the OWC system. In [16], a novel VCSEL array architecture was designed, demonstrating individual beam data rates of 10 Gb/s across a 25 m² indoor area and potential aggregate data rates exceeding 2 Terabit-per-second (Tb/s).

Despite the advantages of VCSELs, one of the significant challenges in OWC systems remains; signal disruption caused by obstacles that block the line-of-sight (LoS) signal between the transmitter and receiver [17]. This is particularly relevant for short-range VCSEL-based OWC, where precise beam steering is crucial. Fortunately, a recent technology, intelligent reflecting surfaces (IRSs), offers a potential solution to address this challenge. IRSs consist of number of low cost, passive and reconfigurable elements that can intelligently manipulate the wireless propagation environment [18], [19]. By creating alternative non-line-of-sight (NLoS) paths towards users, IRSs can circumvent blockages, maintain connectivity, and mitigate their negative effects on overall OWC system performance [20]. By intelligently reflecting signals, IRSs can significantly enhance received signal power and boost data rates. Crucially, unlike active relaying technologies, IRSs achieve this passively, without additional power consumption. This makes them highly attractive for sixth generation (6G) and future wireless networks where spectral and energy efficiency are paramount.

This paper investigates the significant sum rate improvement by integrating IRS in OWC system employing an ADT. Our ADT design utilizes VCSEL array in each branch to precisely direct the light towards both the IRS reflective elements and the users for high-performance data transmission. We evaluate our network model's performance in terms of achievable sum rate. Our results demonstrate that deploying ADT-VCSEL and IRS combination in OWC networks significantly outperforms systems without IRS in terms of achievable sum rate for multiple users.

The rest of this paper is structured as follows: Section 2 outlines the system model setup. Our simulation results and analysis are presented in Section 3 and the conclusions are stated in in Section 4.

## 2. SYSTEM MODEL

In this work, we consider a multiuser downlink indoor IRS-aided OWC network within a room of width × length × height dimensions as shown in Fig. 1. The system has an optical access point (AP) on the ceiling. The AP consists of ADT with multiple branches where each branch consists of N × N array of VCSELs. Each branch is characterized by specific azimuth ($Az$) and elevation ($El$) angles. The values of the $El$ and $AZ$ angles are fine tuned to ensure adequate lighting and to establish high-quality links. By carefully orienting each face with different $Az$ and $El$ angles, the system effectively covers a wide area [21]. Moreover, we assume an IRS mirror array consisting of M × M rotational mirrors, is mounted on one wall of the room. The $K$ users are distributed uniformly on the communication floor and are served by the AP. Each user is equipped with an angle diversity receiver (ADR), which is designed with multiple photodiodes, each photodiode has a narrow field of view (FOV). The ADR design enhances received signal quality and reduces interference. Moreover, the user receives signal from two paths: the direct LoS signal from the AP and through NLoS signal reflected by one of the IRS passive mirrors. Note that, it is important to consider eye safety regulations when using VCSELs for transmission. To ensure eye safety, it is necessary to consider the maximum permissible emission (MPE), which is the maximum irradiance that may be incident upon the eye or skin without causing biological damage [10].

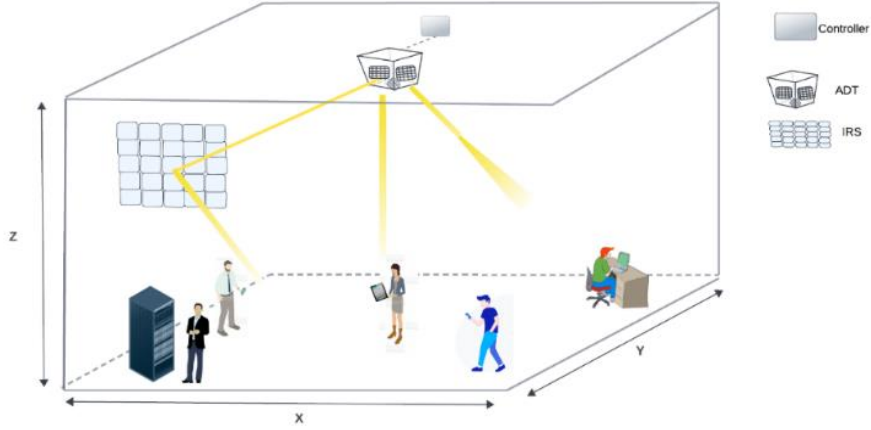

*Figure 1. System model.*

### 2.1 Gaussian beam layout

The distribution of the VCSEL transmitted power depends on three key factors: the beam waist $W_0$, the wavelength $\lambda$, and the distance $d$ between the ceiling and the communication floor. The value of VCSEL beam waist $W_d$ at a user located on the communication floor at distance $d$ is given by [10]

$$W_d = W_0 \left(1 + \left(\frac{\lambda d}{\pi W_0^2}\right)^2\right)^{1/2}, \qquad (1)$$

where $W_0$ is the beam waist at distance 0 from the aperture. Hence, the VCSEL intensity can be calculated based on the radial distance $r$ from the beam centre and the distance $d$. This determines VCSEL spatial intensity distribution across the transverse plane at distance $d$ which can be expressed as

$$I(r,d) = \left(\frac{2 P_t}{\pi W_d^2}\right) exp\left(-\frac{2r^2}{W_d^2}\right), \qquad (2)$$

where $P_t$ is the transmitted optical power of the VCSEL. Consequently, the received power $P_r$ at distance $d$ when the signal passes through an aperture of radius $r_0$ is calculated as

$$P_r = \int_0^{r_0} I(r,d) 2\pi r dr = P_t \left[1 - exp\left(-\frac{2r_0^2}{W_d^2}\right)\right]. \qquad (3)$$

### 2.2 Indoor optical channel model

In our work, the optical channel consists of a LoS signal, representing the direct link between the transmitter and user, and a NLoS component, representing the reflections of the signal by the mirror elements in IRS. Hence the optical channel gain $q_k$ for user $k$ is expressed as

$$q_k = h_{k,l}^{los} + h_{k,m}^{Nlos}, \qquad (4)$$

where $h_{k,l}^{los}$ represents the LoS channel from user AP $l$ to user $k$ while $h_{k,m}^{Nlos}$ is the NLoS reflective signal that reaches user $k$ from mirror $m$ after specular reflection. The total received signal $y_k$ of user $k$ is given by:

$$y_k = q_k\, x + z_k, \tag{5}$$

where $x$ is the transmitted signal from the AP. $z_k$ is the real-valued additive white Gaussian noise (AWGN) with zero mean and variance noise. In this work, the signal-to-interference-plus-noise ratio (SINR) of user $k$ and can be given by

$$\gamma_k = \frac{(R^\circ\, q_k\, P_{tot})^2}{\sigma^2}, \tag{6}$$

where $R$ is the photo detector responsivity, $P_{tot}$ is the transmit power of the VCSEL array and $\sigma^2$ is the noise considered in the channel includes Relative Intensity Noise (RIN), preamplifier noise thermal noise and shot noise. Accordingly, the achievable rate, $R_k$, is calculated by:

$$R_k \cong B \log_2\left(1 + \frac{e}{2\pi}\gamma_k\right). \tag{7}$$

where B is the modulation bandwidth.

## 3. RESULTS

We evaluated the performance of the proposed IRS-aided laser OWC system in an indoor environment with dimensions of 5 m × 5 m × 3 m. The network consists of a single optical AP in the centre of the ceiling. The ADT consists of five branches. One branch is mounted at the center while the rest are placed symmetrically along the four sides. Each branch consists of a 5 × 5 array of VCSELs with 1550 nm wavelength. On the communication floor, four users ($K = 4$) are distributed. Moreover, the area the photodetector and its responsivity are set to $20\ mm^2$ and $0.4\ A/W$, respectively. The rest of the simulation parameters are illustrated in Table I. MATLAB simulation was developed to evaluate the system's performance.

In Fig.2, the achievable sum rates of the system using IRS are compared to a system without IRS. The comparison is based on the sum rates versus the transmit SNR. Moreover, as the number of mirrors in the IRS increases, the achievable rate also increases. The results also show using an IRS with a 10 × 10 mirror array achieves a sum rate of 26% higher than the system with a 5 × 5 array and 71% higher compared to the system without IRS.

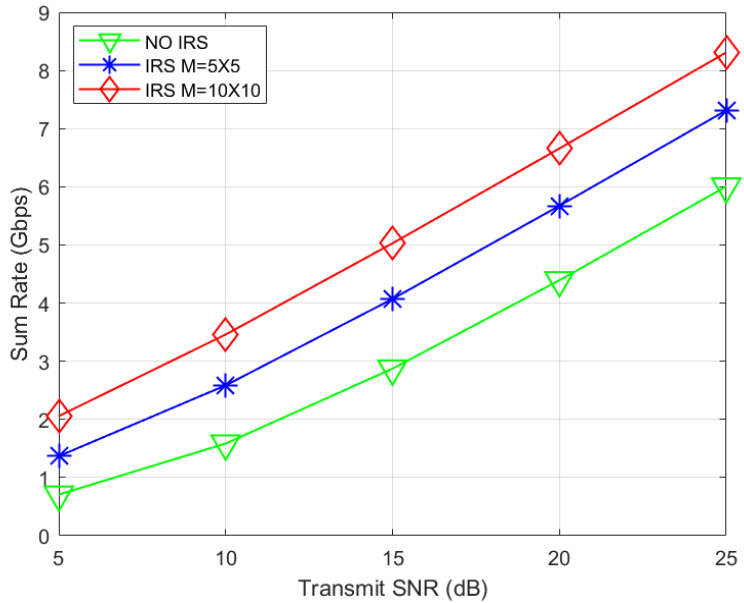

*Figure 2. Sum rate vs transmit SNR.*

The sum rate versus the number of users results is depicted in Fig.3. The results show that the sum rate of integrating an IRS array in OWC network outperforms a traditional OWC network without IRS. Moreover, the results show that the deployment of IRS can significantly boosts the network sum rate because IRS helps enhance the coverage by reflecting the signals towards the users, leading to improve the overall network performance.

Table 1. System parameters [15].

| Parameter | Value | | Parameter | Value |
|---|---|---|---|---|
| Beam waist | 5 $\mu$m | | VCSEL bandwidth | 1.5 GHz |
| TIA noise figure | 5 dB | | Laser RIN noise | -155 dB/Hz |
| IRS mirror elements | 5 x 5 | | Receiver noise current density | 4.47 pA/$\sqrt{Hz}$ |
| Mirror reflectivity | 0.95 | | Mirror area size | 15 cm x 10 cm |
| ADR photodetector branch | 1 | 2 | 3 | 4 |
| Azimuth angels | 0° | 90° | 180° | 270° |
| Elevation angels | 60° | 60° | 60° | 60° |
| Field of View | 25° | 25° | 25° | 25° |

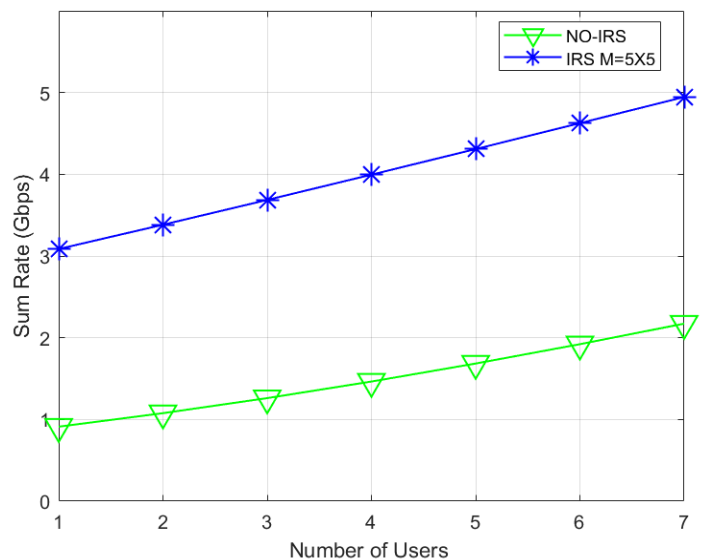

*Figure 3. Sum rate vs number of users.*

## 4. CONCLUSIONS

This paper investigated the potential of integrating VCSELs and IRS to enhance the capacity of indoor optical wireless networks. We proposed a system model that integrated IRS with ADT utilizing VCSELs. Our VCSEL-based system ensures compliance with eye safety regulations while delivering high data rates. Simulation results demonstrate a significant improvement in achievable sum rate with the IRS-aided VCSEL system compared to a traditional OWC network without IRS. Moreover, an IRS with $10 \times 10$ mirror array achieved approximately 26% higher achievable rate compared to a system employing IRS with $5 \times 5$ array and 71% greater gain than a system without IRS. Future research will explore advanced optimization techniques to further boost the overall network performance.

**ACKNOWLEDGEMENTS**
This work has been supported in part by the Engineering and Physical Sciences Research Council (EPSRC), in part by the INTERNET project under Grant EP/H040536/1, and in part by the STAR project under Grant EP/K016873/1 and in part by the TOWS project under Grant EP/S016570/1. All data are provided in full in the results section of this paper.